\documentclass[a4, twocolumn]{article}
\usepackage[utf8]{inputenc}
\usepackage{mathtools}
\usepackage{amsmath,amssymb}
\usepackage{braket}
\usepackage{amsmath}
\usepackage{bm}
\usepackage{authblk}
\usepackage{babel}
\usepackage{graphicx}
\usepackage{placeins}
\usepackage{soul}

\usepackage[right=1.5 cm, left=1.5 cm, top=2cm]{geometry}

\newcommand{\WS}{\textnormal{WS}_2}
\newcommand{\A}{A_{1g}}
\newcommand{\E}{E_{2g}^1}

\title{Fine-Tuning of the Excitonic Response in Monolayer WS$_2$ Domes via Coupled Pressure and Strain Variation}

\date{}
\author[a]{Elena Stellino}
\author[b]{Beatrtice D'Al\`o} 
\author[b]{Elena Blundo}
\author[b]{Paolo Postorino}
\author[b]{Antonio Polimeni}

\affil[a]{\small{Sapienza Universit\`a di Roma, Dipartimento di Scienze di Base e Applicate per l'Ingegneria, I-00185 Roma, Italy}}
\affil[b]{\small{Sapienza Universit\`a di Roma, Dipartimento di Fisica, I-00185 Roma, Italy}}

\begin{document}
\twocolumn[
  \begin{@twocolumnfalse}
    \maketitle
      \noindent  We present a spectroscopic investigation into the vibrational and optoelectronic properties of WS$_2$ domes in the 0-0.65 GPa range. The pressure evolution of the system morphology, deduced by the combined analysis of Raman and photoluminescence spectra, revealed a significant variation in the dome’s aspect ratio. The modification of the dome shape caused major changes in the mechanical properties of the system resulting in a sizable increase of the out-of-plane compressive strain while keeping the in-plane tensile strain unchanged. The variation of the strain gradients drives a non-linear behavior in both the exciton energy and radiative recombination intensity, interpreted as the consequence of a hybridization mechanism between the electronic states of two distinct minima in the conduction band. Our results indicate that pressure and strain can be efficiently combined in low dimensional systems with unconventional morphology to obtain modulations of the electronic band structure not achievable in planar crystals.
      \\\\
  \end{@twocolumnfalse}
       ]      

\section*{Introduction}
Since the discovery of graphene, research on layered materials has progressively grown, making them one of the most studied field in Condensed Matter Physics as promising building blocks for the next generation of electronic devices. In 2D systems, the in-plane confinement limits the available phase space and reduces the electrostatic screening, leading to enhanced carrier correlations and quantum effects. As a result, these crystals exhibit outstanding optical, electronic and magnetic properties, not achievable in their bulk counterparts \cite{Novoselov2016, Bhimanapati2015}. The huge potential of layered materials does not lie only in their intrinsic physical qualities but depends also on the possibility of a fine tuning of their optoelectronic properties. Indeed, due to the atomically thin nature of 2D materials, their properties are very sensitive to perturbations and can be widely modulated through various approaches that include pressure \cite{Stellino2023, Stellino2022}, heterostructuring \cite{Geim2013}, doping \cite{Wang2021}, alloying \cite{Yao2021, Stellino2021}, defects \cite{Jiang2019}, electrical gating \cite{Chen2019}, and strain engineering \cite{Frisenda2017}. The latter, in particular, is recognized as one of the most effective tools to achieve controlled tuning of the structural and optoelectronic properties in low-dimensional crystals \cite{Blundo2021, DiGiorgio2021}, which display much stronger deformation capacity compared to bulk samples.
\\
Differently from 3D materials, in which tensile or compressive forces can be applied directly to the sample, the \textit{all-surface} nature of 2D systems has made it necessary the development of alternative methods for strain modulation \cite{Sun2019}. For instance, low-dimensional flakes can be deposited on top of flexible substrates, which can effectively transfer their deformation to the over-layer \cite{Sun2019, Mohiuddin2009}.  Wrinkles or ripples could also form, either spontaneously or intentionally, when the 2D system is placed onto the substrate \cite{Sun2019} giving rise to composite strain gradients across the sample. 
\\
Two main limitations there exist in strain-applying methods based on layer-substrate systems. The first one is the slippage between the 2D sample and the underlying material, which prevents a full transfer of the strain from the latter to the former \cite{Han2021, Yang2021}. The second one is the sample-substrate interaction \cite{Leonhardt2020}, which considerably affects the optoelectronic properties of 2D crystals. 
\\
In the last years, several works have pointed out the potentialities of domes of 2D crystals as new, intriguing systems for studying the response of monolayers to strain \cite{Blundo_Felici2020, Blundo2020, Blundo_Yildiririm2021}. Domes can be produced on the surface of bulk samples by hydrogen ion irradiation \cite{Cui2023, Dai2018, Tedeschi2019}. They consist of hydrogen-filled, micro-sized, spherical membranes made of a monolayer directly lifted from the underlying bulk lattice. Owing to the membrane curvature, the monolayer undergoes a biaxial tensile strain at the dome centre (resulting in a total strain of about 4 $\%$) and a uniaxial tensile strain in correspondence with the borders (of the order of 2$\%$). Remarkably, in these systems, the dome never experiences an interaction with an external substrate, meaning that: (i) being the strain entirely ascribable to morphological deformation, sliding effects between the monolayer and the underlying crystal are \textit{de facto} prevented, and (ii) the dome does not suffer from the increase in the defect density affecting substrate-supported monolayers. 
\\
In light of these unique assets, the next steps in the study of TMD domes are naturally directed toward the possibility of a controlled modulation of the system morphology and, thus, of the strain applied to the membrane. In this vein, we present, here, the high-pressure investigation of WS$_2$ domes carried out by micro-Raman and micro-photoluminescence (PL) spectroscopy. 
Combining the results from these techniques, we could follow the evolution of the dome shape up to $\sim$0.65 GPa; above this threshold the signal from the monolayer dome was no longer detectable. We found that, on increasing pressure, the system reduces in volume while maintaining a constant basal radius. The pressure-induced distortion of the dome morphology determines an increase in the out-of-plane compressive strain applied at the dome centre, while the in-plane tensile strain remains constant. The pressure evolution of these highly non-uniform strain components is responsible for an anomalous response of the conduction band extremes of the 1L-dome: an hybridization mechanism is proposed between Q and K states, which would result in a non-linear trend of both the exciton energy and radiative recombination intensity. Once the compression-decompression cycle is completed, the original shape of the dome is fully restored, as well as its optical response.

\section{Experimental}
\label{sec:Materials_and_methods}
\begin{figure*}
    \centering
    \includegraphics[width = 0.8 \textwidth]{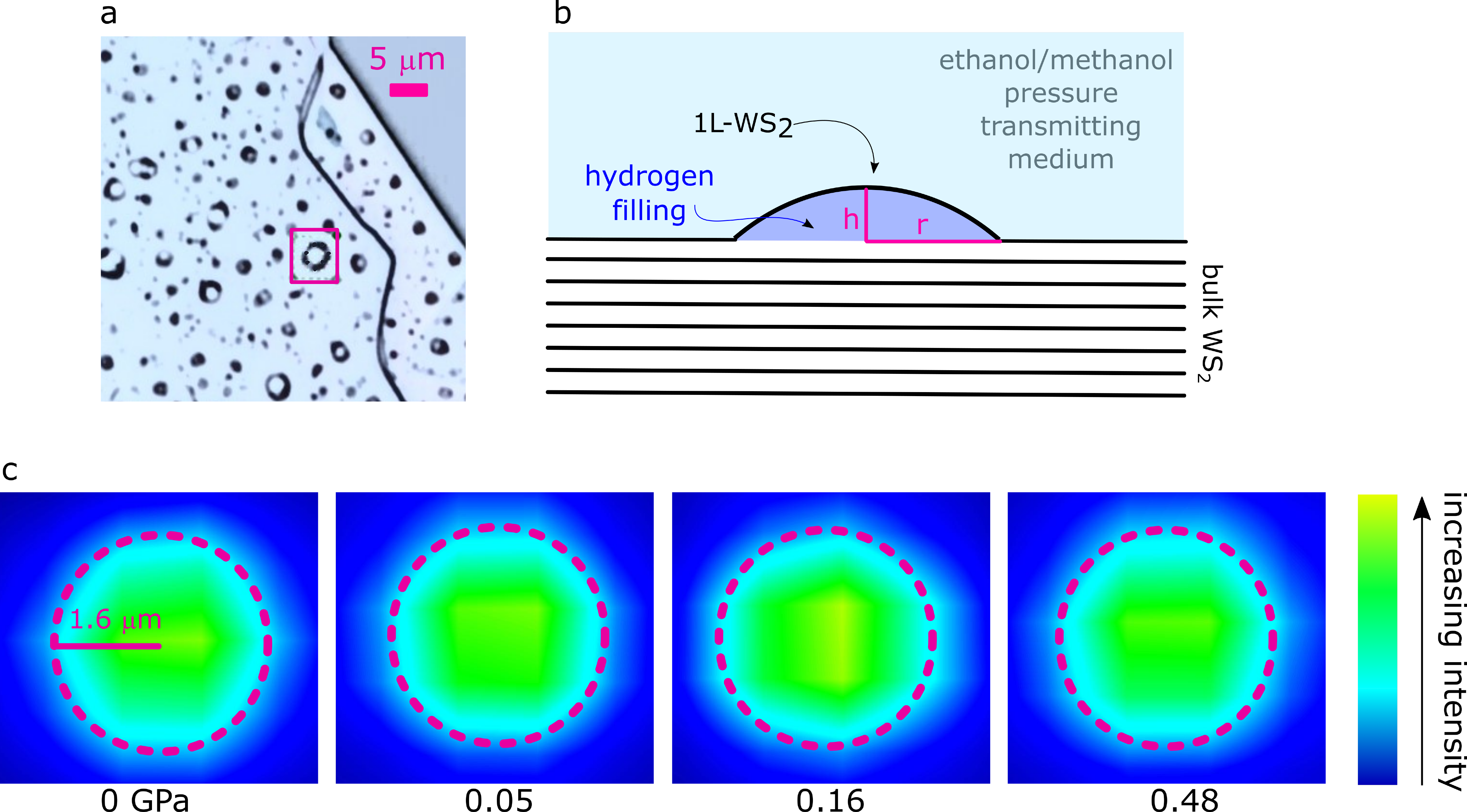}
    \caption{\textbf{(a)}: Microscope image of the investigated $\WS$ sample collected with a 50x objective. The magenta square identifies the dome considered in our measurements. \textbf{(b)}: Schematic representation of a 1L-$\WS$ dome. (c) Color maps obtained at different pressure values collecting PL spectra in a 2D map 6$\times$6 $\mu$m$^2$ with steps of 1 $\mu$m around the dome. The color scale is assigned by integrating over the main PL band in the spectrum.}
    \label{fig:intro_bolla}
\end{figure*}

1L-$\WS$ hydrogen-filled domes were produced by proton irradiation of bulk $\WS$ flakes, as reported in ref. \cite{Tedeschi2019}. As a matter of fact, low-energy (10-30 eV) protons penetrate through the topmost lattice planes (with the sample held at 150 $^{\circ}$C), where they react with ground electrons thus forming H$_2$ molecules. Then, the latter coalesce into highly pressurised micron-/nano-sized volumes finally leading to the formation of H$_2$-filled bubbles with just one monolayer thickness.

Figure \ref{fig:intro_bolla}(a) shows a microscope image of the sample surface we considered in this paper. The surface is studded by round-shaped features with size ranging from several microns to few hundreds of nm, each of these structures being a dome. Data presented in the following refer to one of the two domes measured in our experiment. The dome is schematically represented in figure \ref{fig:intro_bolla}(b) and its curvature makes the 1L-membrane experience, at the centre, a biaxial tensile strain ($\sim$2$\%$) which tends to become uniaxial when approaching the borders \cite{Blundo_Felici2020, Blundo_Yildiririm2021}.
\\
As shown in figure \ref{fig:intro_bolla}(b), the size of the dome is defined by the radius of its circular basis $r$ and by its height $h$. It is well established that, at ambient conditions, the $h/r$ ratio of the domes in van der Waals (vdW) materials is a constant parameter that solely depends on the sample composition \cite{Blundo_Yildiririm2021, Khestanova2016}. In the WS$_{2}$ case, $h/r$ was found to be $0.165 \pm 0.018$ \cite{Blundo_Yildiririm2021}. Considering a   radius $r = 1.6 \,\, \mu$m (see figure \ref{fig:intro_bolla}(a)), we obtain a corresponding height of $0.26 \pm 0.02 \,\, \mu$m for the considered dome. 
\\\\
To carry out high-pressure measurements, a WS$_2$ bulk flake was irradiated by a hydrogen dose equal to about 10$^{16}$ ions/cm$^2$ at 150 $^{\circ}$C leading to the formation of bubbles as displayed in figure \ref{fig:intro_bolla}. The flake was deposited right at the centre of the 600 $\mu$m diameter culet of a screw-driven Diamond Anvil Cell (DAC) by means of a deterministic 2D transfer system. The two diamonds were separated by a 250 $\mu$m thick molybdenum gasket, which had been previously indented down to a thickness of $\sim$30 $\mu$m in correspondence with the culets. A 300 $\mu$m diameter hole was drilled at the centre of the indented gasket by a spark eroder system. In this case, the choice of a Mo gasket and of a relatively large hole, which tends to make the measurement chamber more easily deformable, was necessary in order to obtain a fine increase of the applied pressure in the range 0 to 0.65 GPa. The hole was loaded with a micrometre ruby sphere, exploited as a pressure gauge \cite{Shen2020}, and an ethanol-methanol (4:1) mixture used as pressure transmitting medium (PTM). The latter remains in its liquid state in the whole pressure range here considered, enabling for a fully hydorstatic compression \cite{Klotz2009}.
\\
Micro-Raman and micro-PL spectra were collected using a Horiba LabRam HR Evolution microspectrometer coupled with a grating monochromator (with 600 lines/mm for photoluminescence measurements and 1800 lines/mm for Raman measurements, which give us a spectral resolution of $\sim$3 cm$^{-1}$ and $\sim$1 cm$^{-1}$ respectively) and a CCD detector, with a He-Ne 633 nm laser as a light source \cite{Carpenella2023}. The laser was focused on the sample surface through a 50x objective, resulting in a $\sim 1.5\,\,\mu$m spot size.
\\
At each pressure, we collected a 2D spatial map measuring both Raman and PL spectra in a square 6$\times$6 $\mu$m$^2$ with steps of 1 $\mu$m. The images shown in figure \ref{fig:intro_bolla}(c) are obtained as color maps, in which the color associated with each point corresponds to the PL integrated intensity of the 1L-dome (bearing in mind that 1L crystals show a much higher emission than bulk samples, the PL intensity in the 2D maps marks the dome shape with a high degree of reliability). Based on these plots, we observe a first unexpected result, namely the diameter of the dome remains practically unchanged over the whole pressure run. To analyse Raman and PL peaks, in the following, we have selected at each pressure the spectrum at the centre of the dome in the corresponding color map.
\\\\
High-pressure micro-PL measurements, with the same hydrostatic medium, on a monolayer WS$_2$ directly exfoliated on the diamond culet were also performed as a reference.

\FloatBarrier

\section*{Raman measurements}
\label{sec:Raman_measuerements}

\begin{figure*}
    \centering
    \includegraphics[width = 0.9\textwidth]{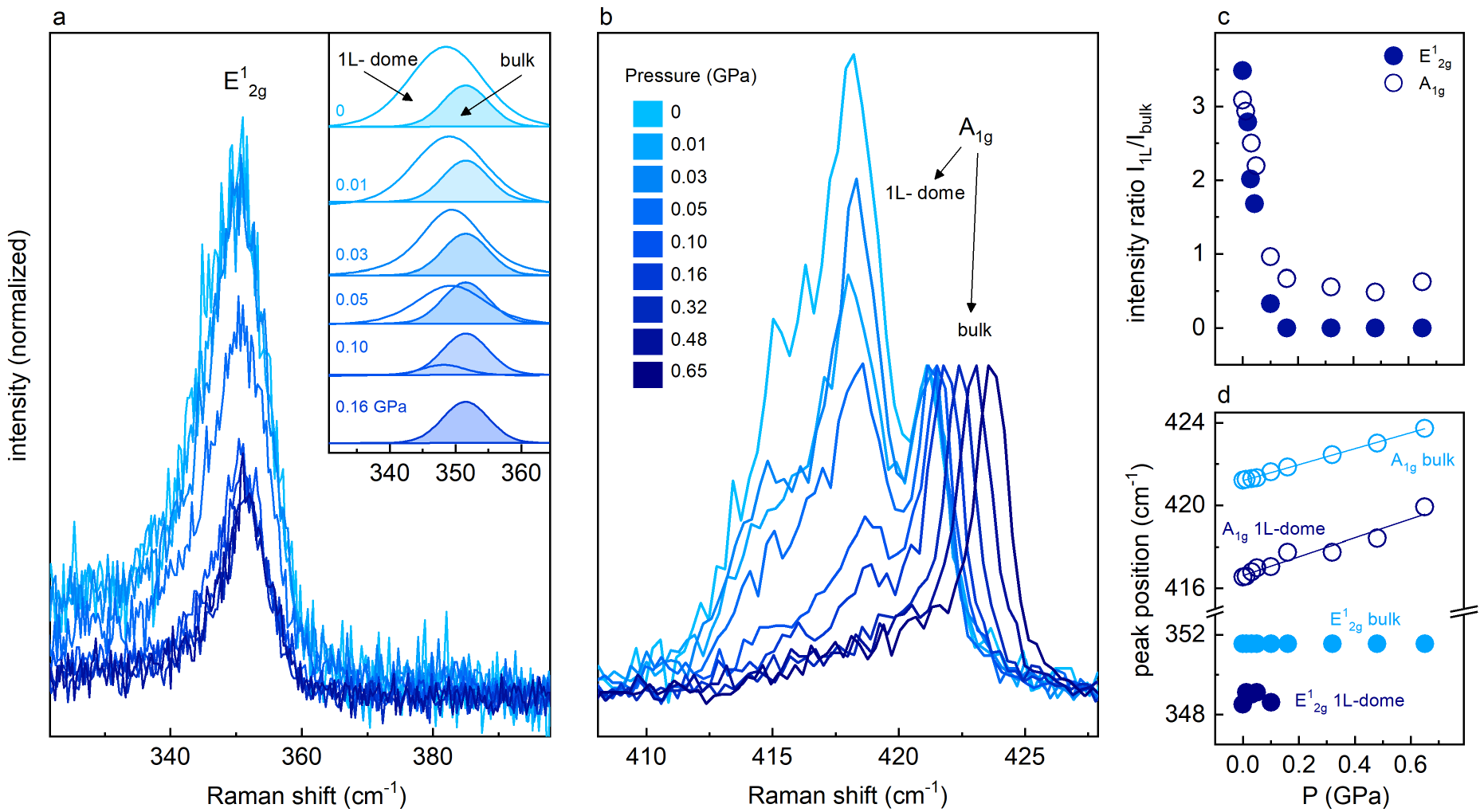}
    \caption{Raman spectra of the 1L-$\WS$ dome highlighted in figure \ref{fig:intro_bolla}(a) at different pressure values. All the spectra were normalized to the maximum of the bulk-$\A$ mode.  \textbf{(a)} Raman spectrum of the in-plane $\E$ mode of the 1L-dome and that of the bulk at increasing pressure values.
    \textbf{(b)} Raman spectrum of the out-of-plane $\A$ mode of bulk and 1L-dome  at increasing pressure values.
    \textbf{(c)} Trend of the ratio between the area of 1L-dome and bulk peaks as a function of pressure for the $\A$ (open dots) and $\E$ (full dots) peaks.
    \textbf{(d)} Pressure trend of the $\A$ (open dots) and $\E$ (full dots) peak centers as a function of pressure for 1L-dome (dark blue) and bulk (light blue). To estimate the frequency of the $\A$ mode of the 1L-dome, at each pressure, we identified the frequency at which the value of the integral curve of the spectrum, in the 410-420 cm$^{-1}$ range, was halved.}
    \label{fig:Raman_measurements}
\end{figure*}

As shown in figure \ref{fig:Raman_measurements}(a),(b), the Raman spectra measured at ambient pressure at the dome centre display low-frequency contributions around 350 cm$^{-1}$ ascribable to the in-plane $\E$ mode \cite{MolinaSnchez2015}, and high-frequency contributions above 410 cm$^{-1}$ ascribable to the out-of-plane $\A$ mode (these modes should be indicated as $E'$ and  $A_1'$ in the $1L$-WS$_2$ crystal \cite{RibeiroSoares2014}, but hereafter we will use the common practice to generalize the nomenclature of the bulk modes to the 1L ones). 
\\
In the spectral feature ascribed to the $\A$ modes in figure \ref{fig:Raman_measurements}(b), we can further distinguish, at ambient pressure, a well-defined peak at $\sim$421 cm$^{-1}$ arising from the $\A$ mode of the bulk substrate and a broader band at $\sim$416 cm$^{-1}$ due to the $\A$ mode of the strained 1L-dome. As for the $\E$ modes, a two-band fit in the range 330-380 cm$^{-1}$ (see inset in figure \ref{fig:Raman_measurements}(a)) allows us to resolve the bulk peak at $\sim$ 352 cm$^{-1}$ and the 1L-dome peak at $\sim$348 cm$^{-1}$. As in the previous case, the 1L-dome contribution is broader compared to that of the bulk substrate possibly due to the non uniform strain of the dome \cite{Blundo2020,Blundo_Yildiririm2021, Tedeschi2019}. All the observed spectral features are in good agreement with those observed in reference \cite{Blundo_Felici2020} collected at the centre of equivalent domes.
\\\\
On increasing pressure, the relative intensity of the $\A$ mode of the 1L-dome rapidly decreases with respect to its bulk counterpart (see figure \ref{fig:Raman_measurements}(c)), while the central frequency of both features linearly blueshifts with comparable pressure rates; see figure \ref{fig:Raman_measurements}(d). A trustworthy analysis of the pressure evolution of the $\E$ modes is partially hampered by the lack of spectral resolution between the 1L-dome and bulk peaks. In this case, we noticed that, above 0.1 GPa, the spectra could be nicely fitted by a single band at $\sim$351 cm$^{-1}$ with constant intensity, ascribable to the bulk contribution. By subtracting this band to the spectra at lower pressure, we could deduce an estimate for the evolution of the 1L-dome peak as a function of pressure. The peak intensity is found to decrease with pressure (see figure \ref{fig:Raman_measurements}(c)), coherently with what observed for the $\A$ case, while the central frequency remains unchanged; see figure \ref{fig:Raman_measurements}(d), similarly to the E$_{2g}$ bulk behavior.
\\\\
The different pressure trend exhibited by the in-plane and out-of-plane phonon frequencies can be ascribed to a different response of the strain components at work in the 1L-dome.
\\
According to ref. \cite{Michail2023} the in-plane $\E$ mode of monolayer WS$_2$ exhibits a redshift rate of $\sim$ -6 cm$^{-1}$/$\%$ with increasing tensile biaxial strain, meaning that even a tiny strain variation causes a well visible shift in the peak center. Therefore, the unresponsiveness we observed for the 1L-dome $\E$ frequency indicates the absence of pressure-induced modifications of the in-plane strain in the dome system. 
\\
On the other hand, the linear blueshift of the 1L-dome $\A$ peak indicates a compression of the membrane in the out-of-plane direction. The overall variation of $\sim$3 cm$^{-1}$ observed for the $\A$ contribution in the 0-0.65 GPa range is coherent with the blueshift rate of $\sim$ 4 cm$^{-1}$/GPa reported in ref. \cite{Kim2017} for planar WS$_2$ monolayers. 
\\\\
As for the peak intensities, knowing that, in the pressure range we measure (0-0.65 GPa), the Raman spectrum of 1L-WS$_2$ is not expected to exhibit appreciable intensity changes, the progressive vanishing of the 1L-dome features has to be ascribed to pressure-induced changes in the 1L membrane morphology. In particular, it seems reasonable to argue that the application of pressure drives a reduction in the dome height $h$, until the distance between 1L crystal and bulk reduces to the sub-nm scale. In this configuration, the vdW interaction between the dome and the underlying bulk increases to the point that the former progressively looses its free-standing monolayer character, leading to a strong quenching of the 1L-dome features in the Raman spectrum.
\\
This hypothesis is strongly supported by the estimates of the dome height $h$ obtained at each pressure assuming that:  (i) the basal radius $r$ remains constant, (ii) the dome can be modelled as a spherical cap, and  (iii) the H$_2$ gas inside the dome volume behaves as an ideal gas. Under these approximations, by the ideal gas law, we find for the height $h$ a non-linear decreasing trend under pressure, which closely resembles those shown in figure \ref{fig:Raman_measurements}(c) for the relative intensity of the 1L-dome features. 
\\\\
Based on these results, we see that on increasing pressure the aspect ratio $h/r$ of the domes is strongly modified, with the basal radius remaining fixed, as witnessed by the spatial maps discussed in the Experimental section, while the height rapidly decreases. Quite remarkably, this abrupt morphological change seems to leave the in-plane tensile strain component unaffected, while the compressive out-of-plane strain steadily increases with pressure. Combining the results obtained for the pressure evolution of morphology and strain in the 1L-dome we see that the Poisson relations governing the mutual dependence between strain and aspect ratio at ambient conditions \cite{Blundo2021} completely fail at describing the dome response at the 10$^{-1}$ GPa scale. In this respect, we underline that the 1L-dome behavior under pressure significantly deviates from its temperature evolution, in which both $r$ and $h$ reduce when decreasing temperature, with the aspect ratio reducing by just $\sim$30 while going from room-temperature to $\sim$33 K (below which H$_2$ liquefies and the dome suddenly deflates) \cite{Cianci2023}.
\\
It is worth noticing that, although in the high-pressure regime the variation in the dome morphology is considerable, once the sample is brought back at ambient conditions, the original Raman spectrum is fully recovered, indicating a complete reversibility of the dome shape, also confirmed by the microscope images collected after the decompression route.

\begin{figure*}
    \centering
    \includegraphics[width= 0.9\textwidth]{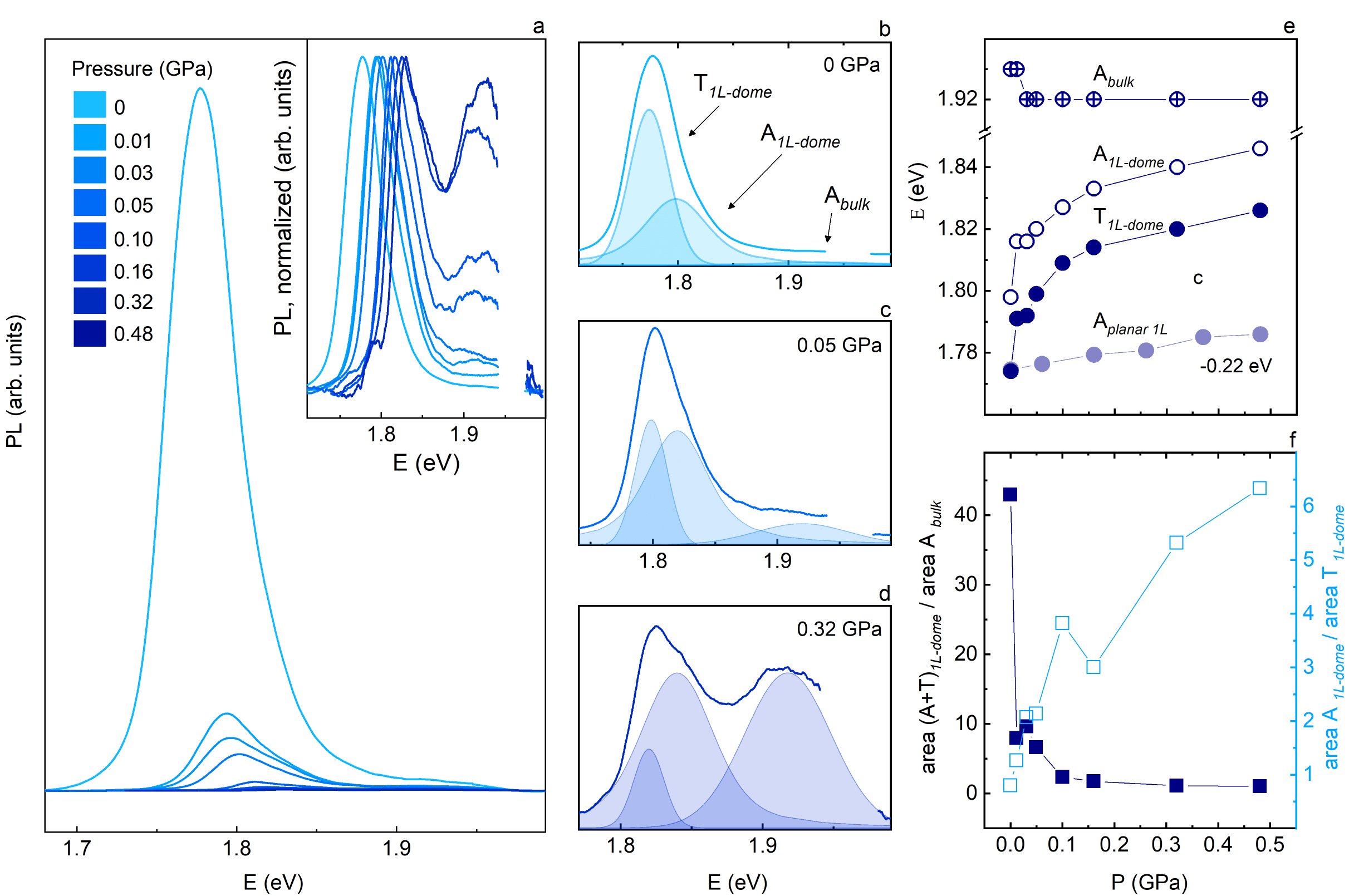}
    \caption{(a) Photoluminescence spectra collected at increasing pressure, on the same dome of the Raman measurements of figure \ref{fig:Raman_measurements}. The inset shows the spectra normalized to the maximum of the low-energy band. Experimental points in the proximity of the Rayleigh peak have been masked for the sake of clarity. (b),(c),(d) Examples of three-peak fitting of the PL spectrum at different pressure values. (e) Pressure trend of the energy position of the different contributions to the PL spectrum: A exciton (A$_{1L-dome}$) and T trion (T$_{1L-dome}$) of the 1L-dome, and A exciton of the bulk substrate (A$_{bulk}$).  The trend of the A exciton of the planar 1L (full dots in light blue) is also reported for sake of comparison. To improve the readability of the plot, the latter trend has been shifted downward by 0.22 eV.
    (f) Pressure evolution of the area of the bands of the 1L-dome (A$_{1L-dome}$ and T$_{1L-dome}$) normalized to the area of the bulk contribution (A$_{bulk}$), full squares in dark blue, and pressure evolution of the area of A$_{1L-dome}$ normalized to the area of T$_{1L-dome}$, open squares in light blue. }
    \label{PL}
\end{figure*}

\section{Photoluminescence measurements}
\begin{figure}
    \centering
    \includegraphics[width= 0.4\textwidth]{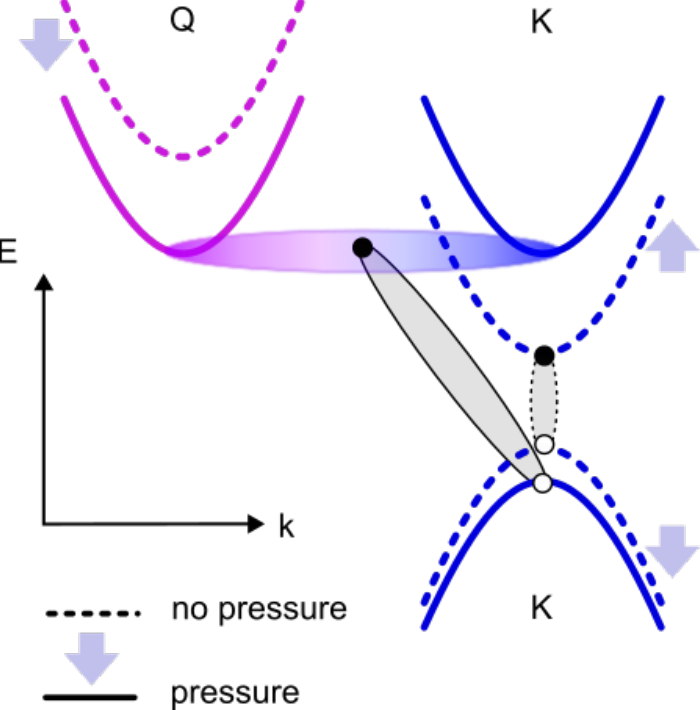}
    \caption{Qualitative representation of the hybridization mechanism described in the text. At ambient conditions (dotted lines), the energy difference between K and Q minima in the CB is too high for the hybridization to take place. On increasing pressure (continuous lines), the recombination process to the VB involves hybridized K-Q states in the CB.}
    \label{fig_final}
\end{figure}

At ambient conditions, the micro-PL spectrum of the system displays a high-intensity contribution below $\sim$1.9 eV and a very low intensity band above 1.9 eV. The former exhibits an asymmetrical profile, which can be accounted for by a two-band fit with centres at $\sim$1.77 and 1.80 eV; see figure \ref{PL}(b). As shown in figures  \ref{PL}(b)-(d) and summarised in figure \ref{PL}(e), on increasing pressure, these two bands blueshift maintaining the same energy distance, while the band above 1.9 eV remains fixed at the same position. We notice a dramatic decrease in the overall PL signal since the first applied pressure (see figure \ref{PL}(a)). At the same time, the relative intensity of the high energy band progressively increases with respect to the low energy contribution; see figure \ref{PL}(b), (c), (d). 
\\
In analogy to the Raman case, once the pressure on the sample is released, the original PL spectrum at ambient conditions is  restored, indicating the complete reversibility of both the dome shape and the optoelectronic properties of the crystal, which confirms the exceptional elasticity and robustness of pressurized domes \cite{DiGiorgio2021}.
\\\\
The specific response of each PL contribution to the applied pressure allows for a reliable assignation of the three bands to distinct radiative recombination mechanisms. 
\\
The high-intensity peaks below 1.9 eV can be ascribed to excitonic recombinations involving carriers at the K point of the first Brillouin zone of the monolayer dome, i.e. the direct exciton (A$_{1L-dome}$) and the  negative trion (T$_{1L-dome}$) at higher and lower energy, respectively \cite{Kesarwani2022, Sebait2021, Wei2016}. On increasing pressure, the blueshift of both A$_{1L-dome}$ and T$_{1L-dome}$ (figure \ref{PL}(e)) indicates a widening of the band gap of the 1L-dome. In this process, the energy separation between trion and exciton, which only accounts for the different electron binding energy in a two- and a three-body system, remains constant as expected \cite{Fu2017}. 
\\
The low-intensity band above 1.9 eV, on the other hand, can be reasonably ascribed to the direct exciton at the K point of the bulk substrate, A$_{bulk}$, which remains practically unaffected by the application of pressure in the 0-0.65 GPa range (figure \ref{PL}(e)). The quencing in the  relative intensity of the 1L-dome features with respect to A$_{bulk}$, shown in figure \ref{PL}(f), is partially ascribable to the pressure-induced reduction in the dome height $h$, in agreement with what was observed for the Raman spectrum, although the rate of reduction of the PL intensity is significantly higher compared to the phonon case.
\\\\
The increase in the 1L-dome band gap, witnessed by the blueshift of both A and T bands, is entirely ascribable to the pressure-induced enhancement of the compressive out-of-plane strain in the monolayer membrane, while the in-plane tensile strain is assumed to remain constant, as discussed in the previous section. Although unaffected by the application of pressure, the in-plane strain gradients at work in the 1L-dome strongly modify the response of its main electronic transitions, determining an evolution for the excitonic contributions which is completely different from that observed in a planar system. As shown in figure \ref{PL}(e), measurements conducted on a WS$_2$ monolayers directly exfoliated on the diamond culet indicate that, in these crystals, the application of pressure up to $\sim$0.5 GPa is responsible for a blueshift of the A exciton (A$_{1L- planar}$) nearly one order of magnitude smaller than that measured for the dome. Moreover, while the rate of increase in the A energy is substantially constant with pressure in the planar case, a strong deviation from linearity is observed in the evolution of both exciton and trion bands in the 1L-dome.
\\
Based on these observations, we hypothesize that, owing to the presence of composite strain gradients, the application of pressure on 1L-dome systems drives an evolution of the electronic band structure which deviates from that predicted for planar monolayers under uniaxial strain variations. In particular, in the present case, the relative conduction band minimum (CBM) at Q (in between K and $\Gamma$) starts matching the energy of the CBM at K at pressures much lower compared to that estimated for planar crystals (i.e. about 2 GPa) \cite{Ma2021}; see figure \ref{fig_final}. This would lead to a hybridization mechanism \cite{Blundo2022, Deilmann2018, Lorchat2021} between the direct recombination process K$_{CB}$-K$_{VB}$ and the indirect Q$_{CB}$-K$_{VB}$ one since the first applied pressure. The mixed direct-indirect nature of the 1L-dome exciton in the high pressure regime accounts for the dramatic intensity drop registered at 0.01 GPa (notice that this value is about 10 times larger than the pressure of the gas inside the dome) a and might be at the origin of the anomalous trend observed for the 1L-dome exciton (and trion) energy in the PL spectrum. The electronic states at K and Q in the CB are known to exhibit an orbital composition that substantially differs in terms of spatial localization, the former hosting mainly in-plane orbitals while the latter shows non-negligible out-of-plane contributions \cite{Fan2015}. As a consequence, the compression of the crystal lattice in the out-of-plane direction drives an increase in the energy of K while the Q extreme is lowered \cite{Ma2021}. The interplay between these opposite trends boils down to the non-linear pressure evolution observed in the mixed direct-indirect exciton: as the contribution from out-of-plane-orbital states to the recombination process increases, the rate of blueshift of both A$_{1L-dome}$ and T$_{1L-dome}$ progressively reduces. As a matter of fact, the Q$_{CB}$ exihibits a slower shift with strain as compared to K$_{CB}$ \cite{Chang2013}.
\\\\
The final remark is on the pressure trend of the relative intensity of A$_{1L-dome}$ with respect to T$_{1L-dome}$, shown in figure \ref{PL}(f). Oppositely from what  reported in high-pressure PL measurements on exfoliated TMD monolayers \cite{Shen2018}, in the present case, we observe a continuous spectral weight transfer from T$_{1L-dome}$ to A$_{1L-dome}$ on increasing pressure. This anomalous behavior can be traced back in the morphological evolution of the 1L-dome. We know from previous works \cite{Kovalchuk2020, Harats2020} that the trion contribution is strongly enhanced when non-uniforms strain components are applied to the system. This is the case of 1L-domes at ambient conditions, in which non-trivial combinations of in-plane/out-of-plane, uniaxial/biaxial strain spatially modulate the monolayer lattice making the trion feature dominate the PL spectrum; see figure \ref{PL}(b). The application of pressure dramatically reduces the aspect ratio of the domes leading to a progressive suppression of the deformation gradients across the system. This favors a more uniform strain distribution, which, in turn, promotes the exciton formation over the trion one. At the same time, it is worth recalling that the 1L-dome, which is basically a suspended membrane, does not suffer from the the pressure-induced enhancement in the sample-substrate interaction, responsible for the pressure-induced increase of the defect-density, and thus of the trion intensity, in exfoliated crystals. 

\FloatBarrier
\section{Conclusions}
The spectroscopic investigation of  WS$_2$ micrometric domes under pressure allowed us to study the system response to compression at the 0.1 GPa scale. 
By combining the results from micro-Raman and mico-PL measurements, we could frame a complete scenario for the pressure evolution of the system, which includes its morphological evolution, the variation of its strain gradients, and the modification of its electronic band structure. 
\\\\
Taking advantage of the well-defined orientation of the lattice displacement in the WS$_2$ phonons, we provided a reliable estimate for the pressure-induced variation of the in-plane and out-of-plane strain components at work on the 1L-dome. Remarkably, the former were found to remain practically unaffected by the great morphological changes undergone by the system, evidencing the need of a radical reformulation of the Poisson relations in the description of the dome response under pressure.
\\
The non-trivial, coupled evolution of morphology and strain in the 1L-dome leads to an anomalous pressure evolution of the band extremes in the CB, responsible for the pronounced non-linearity observed in both the intensity and energy trend of the exciton band under pressure. A hybridization mechanism between distinct CB extremes was proposed at pressure values much lower than those predicted for planar monolayers, resulting in an exciton emission with mixed direct/indirect nature and in-plane/out-of-plane orbital character.
\\\\
Our study demonstrates that external pressure and morphological strain can be efficiently combined in TMD semiconductors to obtain modulations of the electronic band structure not achievable in planar monolayers, defining a new approach to investigate and, possibly, engineer the optoelectronic properties of low dimensional crystals. 
\\
The obtained results point to a scenario in which the tight entanglement of in-plane and out-of-plane strain gradients prevents a straightforward identification of the microscopic processes associated with the evolution of the single components. Although the picture we propose is coherent with the experimental evidence, we believe that an accurate theoretical approach is needed in order to shed light on the mechanisms ruling the system response under pressure.
\\
The robustness of the domes under pressure, the complete reversibility of their optical spectra, and the absence of interaction with external substrates, which considerably affect the PL spectrum of supported monolayers, make this system an ideal playground to push forward the investigation of low-dimensional systems under the simultaneous application of pressure and strain.
\bibliographystyle{unsrt}

\end{document}